\documentstyle[11pt,newpasp,twoside,epsf]{article}
\newcommand {\kms}{\mbox{\,km\,s$^{-1}$}}

\def\edcomment#1{\iffalse\marginpar{\raggedright\sl#1\/}\else\relax\fi}
\reversemarginpar

\begin{document}
\title{Multiple Outflows in AFGL~2688}
 \author{Robert Lucas}
\affil{IRAM, 300 rue de la Piscine, F-38406 Saint Martin d'H\`eres Cedex, France}
\author{Pierre Cox}
\affil{Institut d'Astrophysique Spatiale, B\^at. 121, Universit\'e de Paris XI,
	F-91405 Orsay Cedex, France}
 \author{Patrick J. Huggins}
 \affil{Physics Department, New York University, 4 Washington Place,
         New York, NY 10003, USA}

\begin{abstract}
We present high resolution ($1\farcs 1 \times 0\farcs9$) imaging of the
  proto-planetary nebula AFGL 2688 in the CO (J=2--1) line 
  using the IRAM interferometer. The observations 
  reveal with unprecedented detail the structure and the kinematics of
  the gas ejected by the star over the past few hundred years and 
  exemplify the mechanism by
  which point symmetries are imprinted on the structure of planetary
  nebulae at early stages of their formation.
\end{abstract}

\section{Introduction}
The physical mechanisms which govern the evolution of proto-planetary
  nebulae (PPNe) are poorly documented because relatively few objects are
  known to be in this rapid transition between the AGB and planetary
  nebula phase (Kwok 1993). The role played by high-velocity winds
  which interact with the slowly expanding AGB envelope has been
  recognized as essential in the shaping of the planetary nebulae.
  However, the details of the interaction and the precise evolution
  from the symmetric AGB envelope to the asymmetries which
  characterize planetary nebulae are not well understood.
 
AFGL~2688 (the `Egg nebula') is one of the prime examples of a PPN
  which has evolved from the AGB phase about a hundred years ago (Jura
  \& Kroto 1990). The nearly circular slowly expanding AGB envelope 
  (with a diameter of about $20\arcsec$) is shocked by a warm, 
  optically thin, fast wind (e.g., Young et al. 1992). Studies at high 
  spatial resolution (e.g., Bieging \& Nguyen-Quang-Rieu 1996)
  have shown that the bulk of the 
  molecular gas is concentrated in the central 4$\arcsec$ 
  (or $6 \times 10^{16}$\,cm for an adopted distance of 1~kpc), 
  coincident with the dark lane seen in optical images, with weaker extensions 
  along the north-south axis. Observations at near-infrared wavelengths have 
  revealed that high-velocity H$_2$ gas is present both in the north-south
  and east-west direction suggesting a quadrupolar outflow (e.g., Cox et al. 1997). 
        
This paper presents high angular resolution observations ($1\farcs 07 \times 0\farcs 85$) 
  made in the 
  CO J=2--1 line using the IRAM interferometer which provide a detailed
  picture of the morphology and the kinematics of the gas recently ejected by the 
  central star in AFGL~2688.

\section{Multiple Molecular Outflows}

The results of the observations are 
  summarized in Fig.~1. In the velocity integrated 
  CO map, the distribution of the CO emission consists of a central 
  core $\sim 4\arcsec$ in diameter, with extensions in both the 
  north-south and east-west directions.
  At negative (approaching) velocities, the CO is detected in the
  northern and eastern extensions of the nebula. Along the north
  extension, velocities increase away from the centre; the tip of the
  eastern arm is seen over a range of velocities from $-60$ to $-72$\kms;
  and the highest velocity gas is found near the
  centre of the nebula up to $\sim -80$\kms.  At
  positive (receding) velocities, the CO gas is extended to the south
  and west, with a similar velocity structure to the blue-shifted gas,
  reversed about the systemic velocity.
The channel maps provide definite evidence of two distinct high-velocity 
  outflow directions in AFGL~2688, one along a
  north-south axis at a P.A. of $17\deg$, and the other in a roughly
  orthogonal direction east-west. Detailed examination of the CO data
  reveals that the two main outflows are resolved into a series of
  more collimated, bipolar outflows which are symmetric in direction and
  velocity about the center (Cox et al. 1999) and which are identified
  in Fig.~1: four collimated outflows in the  east-west direction, and 
  three in the north-south direction.  
  It is striking that, for most of these outflows, the tips
  correspond precisely with the H$_2$ peaks seen in the HST image
  (Fig.~1).  The only exceptions are in the central regions and to the
  west side of the equatorial plane, where the high-velocity
  red-shifted CO gas (located behind the dense central gas) has no
  clear H$_2$ counterpart in the HST image, most likely 
  as a result of extinction in the near-infrared.

Along all the outflow axes (A to G), the CO velocity increases with
  distance from the centre. This implies that the observed CO gas is entrained
  by high-velocity jets which could be atomic or ionized. The velocity of the
  CO gas in the outflows is also higher by a factor of $\sim$2.5 than the 
  velocity of the H$_2$ which is close to the expansion velocity of the
  AGB envelope (Kastner et al. 1999). The high-velocity jets 
  entrain the CO gas and shock the molecular hydrogen of the AGB envelope.

Assuming the model-dependent value of $i \sim 16\deg$ for the
  inclination of the north-south axis to the plane of the sky with the
  northern lobe towards the observer (Yusef-Zadeh et
  al. 1984), the deprojected flow velocity of F1 (at
  $6\arcsec$ from the centre) is $22$\kms, corresponding to
  a kinematic age of $\sim 1200$~years. In comparison, the
  projected velocities of the jets D and E are $\sim
  30$\kms, implying ages of $\sim$250 and 125 years, if the
  same inclination is assumed. 
  Finally, we note that the  
  position of the central exciting star of AFGL~2688 recently derived
  from polarisation measurements (J. Kastner, private communication) 
  lies within the positional errors at the intersection of the outflows.

\begin{center}
\plotfiddle{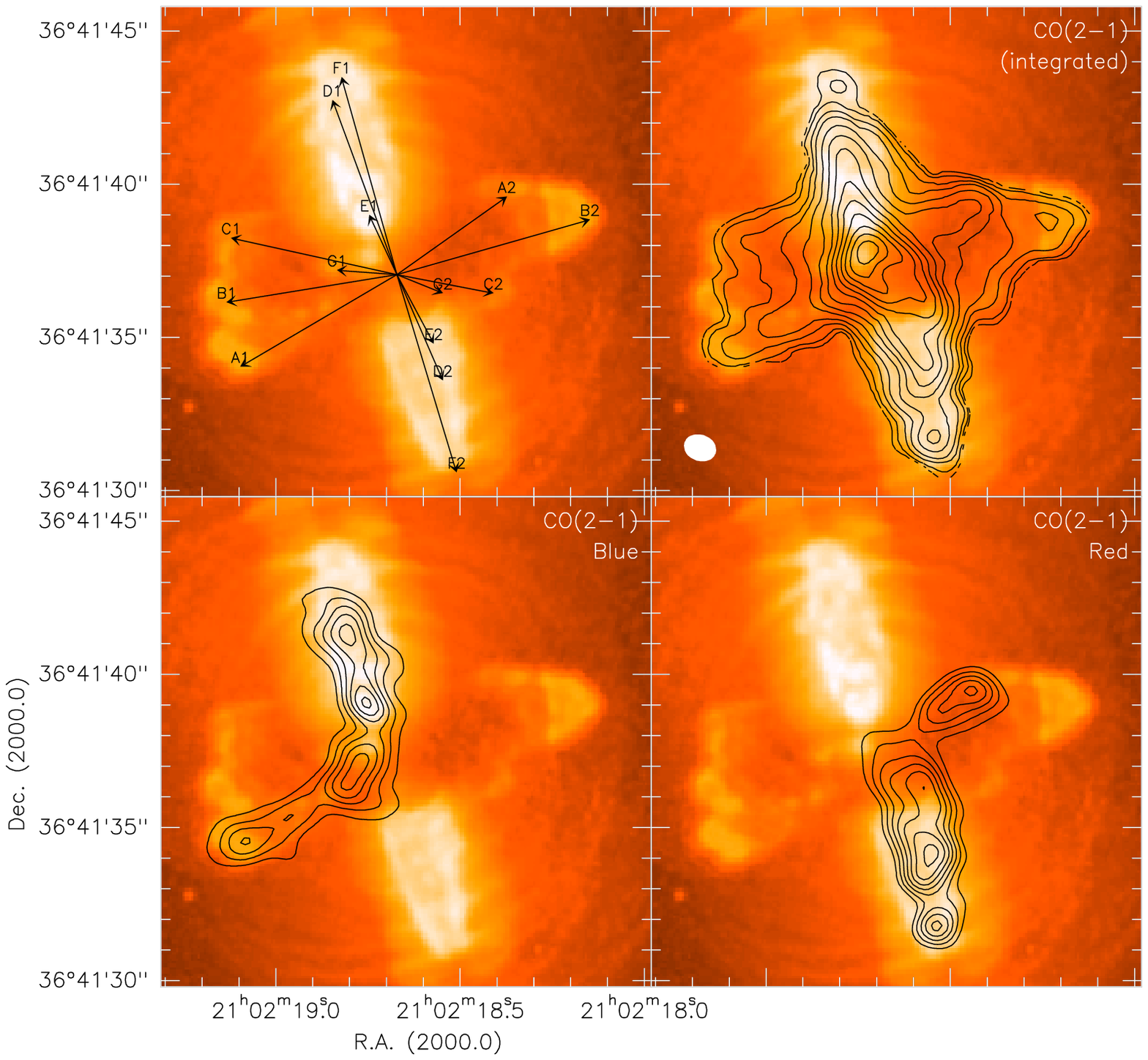}{12cm}{0}{70}{70}{-200}{0}~\\
\parbox{12.2cm}{Figure 1. \hskip 1.5em  
     The Plateau de Bure interferometer data compared to the
     H$_2$ 1--0 S(1) line emission and nearby 2.15~$\mu$m
     continuum from Sahai et al. (1998), in background.  
     Upper right : velocity integrated CO(2--1) with beam size shown in white;
     upper left: outflow axes; 
     lower panels: blue-shifted
     ($-80$ to $\rm -60 \, km s^{-1}$) and red-shifted CO
     ($-22$ to  $\rm -2 \, km s^{-1}$).} \\
\end{center}

\section{The Core Region}

In addition to the collimated outflows, the Plateau de Bure observations
  reveal in the central $2\arcsec$ a shell-like structure
  expanding with a velocity of $\sim 10$\kms (Cox et al. 1999). 
  This CO structure is not aligned 
  with any of the outflow axes and lies at a P.A. of 
  $\sim 54\deg$, which is comparable to that of the 1.3~mm dust
  continuum emission.  The size of this structure
  ($1\arcsec$ or $\rm 1.5 \times 10^{16} \, cm$) 
  implies that it was ejected $\sim$500~years ago and that it could trace the 
  last episode of mass-loss on the AGB;  
  the expansion velocity of this structure is much slower than
  that of the envelope ($\sim 20$\kms - see, e.g., Young et al. 1992).
  Neither the distribution of the dense gas at the
  center nor its kinematics indicate the presence of an equatorial disk
  in AFGL~2688.  The present data thus do not support the interpretation
  of a rotating equatorial disk to explain the east-west kinematics in
  AFGL~2688 (cf.  Bieging \& Nguyen-Quang-Rieu 1996, Kastner et al. 1999).  
  Instead, the CO kinematics reveal the presence of a central expanding
  shell-like structure (not seen in H$_2$) and high-velocity gas with a
  morphology similar to the shocked H$_2$ gas tracing multiple,
  collimated outflows.

\section{Origin and Effects of the Outflows}

The observations discussed here reveal the detailed structure and
  kinematics of the molecular outflows in AFGL 2688. A series of young
  (a few 100~years), high velocity, collimated jets originate from the central star,
  and are directed along the north-south optical axis and in the
  east-west direction. The detailed correlation of the CO outflows with H$_2$
  emission seen in AFGL 2688 provides direct evidence for their
  interaction with the nearly spherical envelope ejected on the AGB.
  The presence of multiple, collimated outflows in two, roughly orthogonal directions
  cannot be explained by the standard two-winds model (e.g., Balick 1987).  
Proto-planetary nebulae with morphologies similar to AFGL~2688 have recently been
found (e.g., Kwok et al. 1998) and 
multipolar jets appear to be a common phenomenon in young planetary
nebulae (Forveille et al. 1998, Sahai \& Trauger 1998). 
Whatever detailed processes are involved, the high-velocity winds must
be intimately linked to the abrupt transition in the evolution of the star
after the AGB phase. One important consequence of the jets is their shaping effect on the
molecular envelope ejected on the AGB. They generate complex point
symmetries in the envelope which later emerge 
in the ionized gas during the PN phase.

\end{document}